\documentstyle[aps,amstex]{revtex}

\newcommand{\ket}[1]{\left| #1 \right\rangle}
\newcommand{\bra}[1]{\left\langle #1 \right|}
\newcommand{\braket}[2]{\left\langle #1 \right| \left. #2 \right\rangle}

\newcommand{\ii}{\'{\i}}

\newcommand{\calD}{{\mathcal{D}}}

\newcommand{\calV}{{\mathcal{V}}}
\newcommand{\nb}{\bar{n}}

\newcommand{\bin}[2]{\left(
                     \begin{array}{c}
                      #1 \\ #2
                     \end{array}
                     \right)}

\begin{document}
\title{Analysing a complementarity experiment on the quantum-classical boundary}
\author{M. O. Terra Cunha}
\address{Departamento de Matem\'atica, Universidade Federal de Minas Gerais, CP 702, Belo Horizonte, 30123-970, Brazil}
\author{M. C. Nemes}
\address{Departamento de F\ii sica, Universidade Federal de Minas Gerais, CP 702, Belo Horizonte, 30123-970, Brazil}
\date{\today}
\maketitle
\begin{abstract}
The complementarity experiment reported in Bertet [{\it{et al.}} (2001), {\it{Nature}} {\bf{411}}, 166.] is discussed. The role played by entanglement in reaching the classical limit is pointed out. Dissipative and thermal effects of the cavity are calculated and a simple modification of the experiment is proposed in order to observe the progressive loss of the capacity of ``quantum erasing''as a manifestation of the classical limit of quantum mechanics.
\end{abstract}

Since the early days of Quantum Theory, the differences between the so called {\emph{quantum world}} and {\emph{classical world}} have been noted and the problem of {\emph{the classical limit of quantum mechanics}} was established. The Einstein-Bohr debate \cite{EB}, the EPR ``paradox'' \cite{EPR35}, and Schr\"odinger's cat \cite{Sch35} are some of the first examples of the difference between quantum and classical. The so called {\emph{hidden variable theories}} form one alternative to make the classical limit smooth \cite{Bel73}, however Bell inequalities impose restrictions on certain hidden variable theories, which some quantum states do violate \cite{SpU}. During the evolution of quantum theory, {\emph{entanglement}} has been recognized as the essential element: present in the states which violate Bell inequalities, consequently at the root of the distinction between {\emph{quantum}} and {\emph{classical}} worlds, entanglement is generally created by the interaction between two initially independent subsystems. The unavoidable interaction between a physical system and the degrees of freedom of its environment tends to generate entanglement, which manifests itself as loss of purity of the quantum state of the system of interest, in a {\emph{decoherence process}} \cite{Giuetal}. For these reasons, entanglement plays this double role: on one side marking the difference between quantum and classical, on the other making it possible that this boundary be smoothly crossed  \cite{Zur91}. In the last decades, all this discussion has been accompanied by experiments which investigate the so called mesoscopic regime, the boundary between the microscopic quantum world and the macroscopic classical world \cite{exp}. In this Letter we will discuss one of these experiments \cite{Beretal01}. 

In the cited work \cite{Beretal01}, the authors show a pair of variations on the Ramsey interferometer, which are compared to Mach-Zehnder interferometers with one or both beam splitters passing from quantum to classical domain. In Ramsey interferometry, a two level atom initially in an eigenstate $\ket{e}$ passes through a {\emph{Ramsey zone}}, {\it{i.e.}}: a region in which the atom interacts with a {\emph{classical}} field, in which it is taken to be in a superposition state like $\ket{\psi} = \frac1{\sqrt{2}}\left( \ket{e} + \ket{g}\right)$. As $\ket{e}$ and $\ket{g}$ correspond to different eigenenergies, the temporal evolution of this state imposes a relative phase between this two state components: $\ket{\psi \left( t\right)} = \frac1{\sqrt{2}}\left( \ket{e} + e^{-i\omega t}\ket{g} \right)$. Another Ramsey zone recombine the two components in a state described by $\ket{\psi '\left( \phi \right)} = i\sin \left( \frac{\phi}2 \right) \ket{e} + \cos \left( \frac{\phi}2 \right) \ket{g}$, where $\phi$ is the relative phase accumulated by the two state components between the two Ramsey zones. The analogy with the Mach-Zehnder interferometer is perfect: the Ramsey zones work as beam splitters and the $\phi$ phase difference corresponds to the optical path difference (divided by $\frac{\lambda}{2\pi}$). With this idealized description, the variation of $\phi$ implies interferometric fringe patterns like $\cos ^2\left( \frac{\phi}2\right)$ on the observation of the probability of detecting an atom in the state $\ket{g}$. In the experiment, the ``classical limit'' is approached in two distinct ways, with two different apparatus schemes. When only one ``beam splitter'' is varied from quantum to classical, the classical limit is attained through the mean photon number in the cavity mode used as Ramsey zone, in an example where the classical world is reached as the ``large quantum numbers limit''; in case both ``beam splitters'' are considered, a different ``classical limit'' can be attained by a decoherence process.  We will discuss a little bit further each case, stressing the different roles played by entanglement in each one. Anticipating the conclusion, in the first case the classical limit corresponds to the evanescence of atom-field entanglement, while in the second it is owed to the creation of field-environment entanglement.

In the first scheme, one mode of a Fabry-Perot resonator $C$ (photon damping time $T_{cav} = 1ms$, with $\nu = 51.1 GHz$) is used to create the first Ramsey zone. This mode is previously fed by a pulsed microwave generator, which creates a coherent state \cite{Gla} of complex amplitude $\alpha$, $\ket{\alpha} = \sum _n c_n \ket{n}$, where $c_n = \exp \left( -\left| \alpha \right|^2 / 2\right) \alpha ^n / \sqrt{n!}$, and the mean photon number is given by $N = \left| \alpha \right|^2$. The idealized description ({\it i.e.} given by the Jaynes-Cummings model) of the resonant interaction of an atom initially in the state $\ket{e}$ with the field mode initially in the coherent state $\ket{\alpha}$ is given by
\begin{equation}
\ket{e} \otimes \ket{\alpha} \longmapsto  \ket{e} \otimes \ket{\alpha _e} + \ket{g} \otimes \ket{\alpha _g} ,
\label{pulso}
\end{equation}
where \cite{note}
\begin{equation}
\begin{array}{ccl}
\ket{\alpha _e} &=& e^{-i\nu t} \sum _n e^{-in\nu t} c_n \cos \left( \Omega _n t_{\alpha}\right) \ket{n}, \\
\ket{\alpha _g} &=& -ie^{-i\nu t} \sum _n e^{-in\nu t} c_n \sin \left( \Omega _n t_{\alpha}\right) \ket{n+1},
\end{array}
\end{equation}
with $\Omega _n = \Omega \sqrt{n+1}$, where $\Omega$ is vacuum Rabi frequency, and $t_{\alpha}$ is given by the condition 
\begin{equation}
\braket{\alpha _e}{\alpha _e} = \braket{\alpha _g}{\alpha _g} = \frac12,
\label{cond}
\end{equation}
which is the $\frac{\pi}2$ pulse condition, analogous to the $50-50$ beam splitters condition on the Mach-Zehnder interferometer. By defining two new states
 $\ket{\alpha _{\pm}} = \frac12 \left( \ket{\alpha _e} \pm \ket{\alpha _g} \right)$, we can rewrite the state in equation (\ref{pulso}) as
\begin{equation}
\ket{e} \otimes \ket{\alpha _e} + \ket{g} \otimes \ket{\alpha _g} = 
\left( \ket{e} + \ket{g} \right) \otimes \ket{\alpha _+} + \left( \ket{e} - \ket{g} \right) \otimes \ket{\alpha _-}.
\label{emaranha}
\end{equation}
It is easy to identify the degree of entanglement at the state given by equation (\ref{emaranha}), provided by the norms of the state vectors $\ket{\alpha _{\pm}}$: if $\braket{\alpha _+}{\alpha _+} \approx \braket{\alpha _-}{\alpha _-}$ then there is much entanglement in the state, while $\braket{\alpha _-}{\alpha _-} \approx 0$ (resp. $\braket{\alpha _+}{\alpha _+} \approx 0$) implies that the state is very well approximated by the factorized state $\left( \ket{e} + \ket{g} \right) \otimes \ket{\alpha _+}$ (resp. $\left( \ket{e} - \ket{g} \right) \otimes \ket{\alpha _-}$).

The presence or absence of entanglement can be interpreted as {\emph{information}} in the context of {\emph{``which way'' interferometers}} \cite{WW}. The condition $\braket{\alpha _-}{\alpha _-} \approx 0$ means that the states $\ket{\alpha _e}$ and $\ket{\alpha _g}$ are very close, and in this case the field mode does not carry (a large amount of) information about the atomic state. On the contrary, if $\braket{\alpha _+}{\alpha _+} \approx \braket{\alpha _-}{\alpha _-}$ then the states $\ket{\alpha _e}$ and $\ket{\alpha _g}$ are almost orthogonal to each other, which implies that atom and field share an EPR like correlation \cite{EPR35}: knowledge on the field state implies knowledge on atomic state and vice versa. Going a little further on the Mach-Zehnder analogy, the same analysis describes the case where  the first beam splitter of the apparatus is attached  to a variable mass pendulum (see figure 1). In the limit case where this mass is very tiny, the momentum kick given by the reflected photon would made the movement states of the pendulum distinguishable for photon reflection and transmission, what can be interpreted as the existence of ``which (photon) path'' information, or more precisely, a large degree of entanglement between the photon and the pendulum. However, in the case of macroscopic pendulum mass, its movement state is practically insensitive to the photon kick, there being no entanglement or which path information.

Back to the experiment, as in the next steps only the atom will be involved, we can eliminate the field degree of freedom by taking the partial trace of the density operator which describes atom-field system. The atomic state is then described by
\begin{equation}
\rho _{at} = \left[
\begin{array}{cc}
\braket{\alpha _g}{\alpha _g} &  \braket{\alpha _e}{\alpha _g} \\
\braket{\alpha _g}{\alpha _e} &  \braket{\alpha _e}{\alpha _e}
\end{array}
\right] = \left[
\begin{array}{cc}
\frac12 &  \braket{\alpha _e}{\alpha _g} \\
\braket{\alpha _g}{\alpha _e} &  \frac12
\end{array}
\right] ,
\end{equation}
where the condition (\ref{cond}) was used. As the $\phi$ phase is accumulated, the atomic state will become
\begin{equation}
\rho _{at} = \left[
\begin{array}{cc}
\frac12 & e^{i\phi} \braket{\alpha _e}{\alpha _g} \\
e^{-i\phi} \braket{\alpha _g}{\alpha _e} &  \frac12
\end{array}
\right] .
\end{equation}
After this, the atom passes through a second Ramsey zone, where it interacts with a short lived field mode (photon decay time lower than $10 ns$), which is sustained by a microwave generator. This kind of Ramsey zone was described in reference \cite{Kimetal99}, where classical behaviour is obtained with low quantum numbers given by the drastic reduction of atom-filed entanglement caused by field dissipation. A pictorial description is that the atom does not have time to entangle with the photon because ``it decays first'', and the interaction is just with an average field, instead of an individual photon. By adjusting the interaction time for another $\frac{\pi}2$ pulse, the atomic state evolves to
\begin{equation}
\rho _{at} = \left[
\begin{array}{cc}
\frac12 + \Re \left( e^{i\phi} \braket{\alpha _e}{\alpha _g} \right) &
i \Im \left( e^{i\phi} \braket{\alpha _e}{\alpha _g} \right) \\
-i \Im \left( e^{i\phi} \braket{\alpha _e}{\alpha _g} \right) &
\frac12 - \Re \left( e^{i\phi} \braket{\alpha _e}{\alpha _g} \right)
\end{array}
\right] .
\label{rhoat}
\end{equation}
By varying the $\phi$ phase, one can obtain an interference fringe pattern on the probability of detecting an atom at the $\ket{g}$ state. In the referred work \cite{Beretal01}, fringe patterns are shown for different mean photon number ({\it i.e.} different values of $\left| \alpha \right| ^2$). Clearly, the fringe contrast grows with $N$. To treat this contrast in a quantitative way, one can use the pattern {\emph{visibility}}, which is defined for a pattern of intensity $I$ by
\begin{equation}
\calV = \frac{I_{max}-I_{min}}{I_{max}+I_{min}}.
\label{vis}
\end{equation} 

Using the state described in equation (\ref{rhoat}), we get $\calV = 2\left| \braket{\alpha _e}{\alpha _g}\right|$. But several effects tend to diminish this fringe contrast, among them we can cite: residual thermal effects, which forbid the field to be taken initially to vacuum state, consequently forbidding the creation of perfect coherent states; detector imperfections, in which the ionization curves of $\ket{g}$ and $\ket{e}$ states have a small overlap, which allow some false readings; dispersion in velocity and position of the atomic beam and imperfections in the ``interaction switches'', which lead to imperfections on the $\frac{\pi}2$ pulses. The authors made a one parameter fit of the data to the discussed theoretical curve and obtain the fitting constant $\eta = 0.75$. This value of $\eta$ can be interpreted as the larger visibility that could be attained with all the imperfections inherent to the experimental setup. This value will be used as a parameter in further discussions.

In the next setup, the same cavity field mode is used to generate both ``Ramsey zones''. The atom, initially in state $\ket{e}$, interacts with the field mode for the time necessary for a $\frac{\pi}2$ pulse. By the Stark effect, the atomic transition is put out of resonance, and the system evolves without interaction for a time $\tau$ (in the experiment, $\tau = 16 \mu s$). Then, the electric field is again tuned to resonance, and another $\frac{\pi}2$ pulse  interaction takes place. Finally, the atom leaves the cavity and goes to the detection region. The Mach-Zehnder analog is to consider both beam splitters attached to the same pendulum (see figure 2). The intense field case, analogous to massive pendulum, will not be explored, because classical limit were already attained, and the only expected effect due to the use of the same field mode is a little attenuation caused by dissipation. In the ``quantum'' case, on the contrary, the entanglement created in the first interaction will play an important role on the second pulse. We will first make an idealized description of the system evolution ({\it i.e.} by considering the atom-field system isolated), and then we will consider the important effects of cavity imperfections, by coupling the field mode to a thermal reservoir.

For simplicity and to compare to the experiment, we will consider the initial field state to be the vacuum. As in equation (\ref{pulso}), the first ``Ramsey zone'' dynamics, with field initial state as vacuum, can be described by
\begin{equation}
\ket{e} \otimes \ket{0} \longmapsto \frac1{\sqrt{2}} \left\{ \ket{e} \otimes \ket{0} -i \ket{g} \otimes \ket{1} \right\}.
\end{equation}
With the Stark effect accumulated phase, the state evolves to
\begin{equation}
\ket{\Psi} = \frac1{\sqrt{2}} \left\{ \ket{e} \otimes \ket{0} + e^{i\phi} \ket{g} \otimes \ket{1} \right\}.
\label{Psi}
\end{equation}
As discussed in the first setup, the above state does not exhibit interference fringes. This can be interpreted as the existence of atomic state information on the field state. But now the same field which is entangled with the atom will be used for the second pulse. This interaction can be described by
\begin{equation}
\ket{\Psi} \longmapsto -e^{i\frac{\phi}2} \left\{ i \sin \frac{\phi}2 \ket{e} \otimes \ket{0} - \cos \frac{\phi}2 \ket{g} \otimes \ket{1} \right\},
\label{final2}
\end{equation}
where it is made clear that the probability of detecting an atom in $\ket{g}$ state reexhibits interference fringes. There is a strong parallel with the so called {\emph{quantum eraser}} \cite{apagador}, where the ``which way'' discriminator state is manipulated in order to exhibit interference fringes in the coincidence detection counts of the interferometric particle ({\it{e.g.}}: the atom) and the ``which way'' discriminator ({\it{e.g.}}: the field). This can be considered as a {\emph{conditional}} interference pattern: {\emph{if}} the discriminator is detected in some state, {\emph{then}} a pattern is obtained in the interferometric particle detections; in another way, interference patterns are obtained only if both degrees of freedom are detected {\emph{in coincidence}}. The phenomenon here is different: there is no need of coincidence detection; the single atom counts do exhibit interference fringes, in an effect called {\emph{unconditional quantum erasure}}. One should note that the state given by equation (\ref{final2}) does exhibit entanglement (except for special $\phi$ values), and fringe patterns are observed just because this signal is already coded at the two occupied level populations ({\it{i.e.}} $\ket{e} \otimes \ket{0}$ and $\ket{g} \otimes \ket{1}$). Any kind of attempt to determine a relative phase signal between these two state components by using only one part of the system (the atom {\emph{or}} the field) will be subjected to $\phi -$value dependent restrictions imposed by entanglement. The imperfections present in the first setup should also manifestate themselves in this second one, and even the imperfections in the ``interaction switches'' must be more effective, as they are used also for the second pulse, so, even with this idealized description, we must expect the interference pattern's visibilities not to exceed the $\eta = 75\%$ value.

It can be argued that, in a certain sense, any interferometer is a realization of an unconditional ``quantum'' eraser. A classical Mach-Zehnder interferometer is again a good example. After the first beam splitter, the two paths are distinguishable, and the wave vector $\vec{k}$ is a good {\emph{which path}} discriminator. No interference pattern can be observed with a detector in one of the arms in between the two beam splitters. To keep this discussion on, consider an incomplete Mach-Zehnder, where we put a detector in place of the second beam splitter. If the detector is aligned with one of the arms, only light from this arm is recorded, and no interference is obtained (figure 3{\bf a}). But if the detector is balanced in a way that light from both arms reach the detector, an interference pattern can be formed by varying the detector's position (figure 3{\bf b}). This situation can be compared to the ``quantum'' eraser: by selecting a linear combination of the {\emph{path states}} of the discriminator, one obtains a fringe pattern. To put the second beam splitter instead of a detector can be viewed as a way of ``rotating the basis'' of the discriminator, allowing fringe (and ``anti-fringe'') patterns in both arms of the interferometer. In this sense, the Mach-Zehnder interferometer (and analogously, any interferometer) can be viewed as a ``quantum'' eraser. One can consider it as an unconditional eraser only to the extent that all the alignment and balance work necessary to the interferometric scheme  already guaranteed all the necessary coincidences for the patterns to be observed, {\it{i.e.}}: to make the interferometric alternatives indistinguishable. No quantum physics is needed in this discussion! Quantum physics appear only when we realize that this can also be done with matter waves, as in Ramsey interferometry.

Now let us take into consideration the finite lifetime of the photon. To this end, we will consider the cavity field mode, with creation (annihilation) operator $a^{\dagger}$ ($a$), coupled to a thermal reservoir. This interaction can be described by the master equation \cite{mest}
\begin{equation}
\frac{d}{dt}\rho _f = -i\left[ H_f,\rho _f \right] + \calD \rho _f ,
\label{master}
\end{equation}
where $H_f$ is the field mode Hamiltonian ($\hbar = 1$), $H_f = \nu \left( a^{\dagger}a + \frac12 \right)$, $\rho _f$ is the field mode density operator and $\calD$ is the {\emph{dissipator}}, which takes care of the nonunitarity of the temporal evolution, due to the interaction of the field with the reservoir \cite{Lin}. We will work with the usual dissipator, which is adequate for the experimental regime in question, given by
\begin{equation}
\calD \rho _f = k\left( \nb +1\right) \left\{ 2a\rho _fa^{\dagger} - a^{\dagger}a\rho _f - \rho _fa^{\dagger}a\right\} + k\nb \left\{ 2a^{\dagger}\rho _fa - aa^{\dagger}\rho _f - \rho _faa^{\dagger}\right\} ,
\label{dis}
\end{equation}
where $\nb$ is the mean photon number of the heat bath with energy $\nu$ and $k = \left( 2T_{cav}\right)^{-1}$ is the damping constant. We should note that the first term is present even at null temperature ($\nb = 0$), while the other is a purely thermal effect. It is easy to solve equation (\ref{master}) in the case $\nb = 0$, and the result also has a simple interpretation: if $\nb = 0$, then the reservoir thermal state is the vacuum state. Consequently, the atom-field system can only lose excitations to the reservoir. The non-interacting atom-field system Hamiltonian spectrum,
\begin{equation}
H = \omega \left( \ket{e}\bra{e} - \ket{g}\bra{g}\right) + H_f,
\end{equation}
has ground state $\ket{g}\otimes \ket{0}$ and equally spaced almost degenerated doublets given by $\ket{g} \otimes \ket{n+1}$ and $\ket{e} \otimes \ket{n}$. The initial state given by equation (\ref{Psi}) is generated by the first doublet. As this system cannot gain excitations, all upper levels will remain unoccupied, and the system will effectively behave as if its space state were $3$-dimensional. The system state cannot be described by a state vector, due to the non-unitary evolution, but its density operator at a time $\tau$ after the first pulse will be given by
\begin{equation}
\rho \left( \tau \right) = \frac12 \left[
\begin{array}{cccc}
1-e^{-2k\tau} &&& \hspace{1cm} \\
& e^{-2k\tau} &e^{i\phi}e^{-k\tau} \\
& e^{-i\phi}e^{-k\tau} & 1 \\ &&&
\end{array}
 \right].
\label{rhotau}
\end{equation}
This result can be easily interpreted: the population of the $\ket{g}\otimes \ket{1}$ component exponentially decays to $\ket{g} \otimes \ket{0}$. This fed component does not have any degree of coherence with the others, while the remaining part keeps the maximum degree of coherence that it can support with the population of the $\ket{e} \otimes \ket{0}$ level. A new $\frac{\pi}2$ pulse is applied and the system can now be described by
\begin{equation}
\rho _{\frac{\pi}2}\left( \tau \right) = \frac12 \left[
\begin{array}{cccc}
1-e^{-2k\tau} &&&  \hspace{1cm}\\
& \frac12 \left( 1+e^{-2k\tau}\right) + e^{-k\tau} \sin \phi &
e^{-k\tau} \cos \phi +\frac{i}2 \left( 1 - e^{-2k\tau} \right) \\
& e^{-k\tau} \cos \phi -\frac{i}2 \left( 1 - e^{-2k\tau} \right)
& \frac12 \left( 1+e^{-2k\tau}\right) - e^{-k\tau} \sin \phi \\ &&&
\end{array}
 \right].
\end{equation}
With the above described state it is immediate to calculate the probability to obtain an atom in the $\ket{g}$ state by summing the two present channels:
\begin{equation}
P_g = \frac14 - \frac14 e^{-2k\tau} + \frac12 e^{-k\tau}\sin \phi ,
\label{pg0}
\end{equation}
which represents an interference pattern with visibility
\begin{equation}
\calV = \frac{2 e^{-k\tau}}{3 - e^{-k\tau}}.
\label{vis0}
\end{equation}
The plot of this function is represented by the upper curve in figure 4. With the data from the Bertet {\it et al.} experiment \cite{Beretal01}, $T = k\tau = .008$, this analysis gives $\calV = 98.8\%$. By normalizing to the maximum visibility given by the first setup, one obtains $\calV = 74.1\%$.

For non null temperature, the master equation (\ref{master}) can also be solved, but the solution will have all doublets occupied. We have used a method which appeared in reference \cite{Wanetal01}, which is based on the linearity of equation (\ref{master}) with respect to $\rho _f$, and keeps some resemblance with the one adopted in reference \cite{met}. The detailed calculation can be found in \cite{det}. Using the initial state ($\ref{Psi}$), evolving the non interacting atom-field system by the equation (\ref{master}) for a time $\tau$, and then using Jaynes-Cummings Hamiltonian to evolve the system for a time $\chi$, we get the atomic state (after field partial trace):
\begin{equation}
\rho _{at} = P_g \ket{g}\bra{g} + P_e \ket{e}\bra{e},
\end{equation}
with $P_g + P_e = 1$, and
\begin{equation}
P_g = P_g^c + P_g^o \sin \phi ,
\label{pad} 
\end{equation}
where the constant term $P_g^c$ is given by the sum of four parts:
\begin{eqnarray}
P_g^{c1} &=& \frac12 \sum _{j=0}^{\infty} e^{-2jT } \frac{\nb ^j - j\nb^{j-1}}{\left( 1+\nb \right) ^{j+1}} ,\\
P_g^{c2} &=& \frac12 \sum _{j=0}^{\infty} e^{-2jT } \frac{\nb ^j - j\nb^{j-1}}{\left( 1+\nb \right) ^{j+1}} \sum _{m=0}^{j-1} \left[ -\left( 1+\nb \right) \right] ^{-m-1} \bin{j}{m+1} \cos ^2\left( \Omega _m \chi \right) ,\\
P_g^{c3} &=& \frac12 \sum _{j=0}^{\infty} e^{-2jT } \frac{\nb ^j}{\left( 1+ \nb \right)^{j+1}} \sum _{l=0}^j \left( -\nb\right) ^{-l} \bin{j}{l} \sum _{m=l}^{\infty} \left( \frac{\nb}{1+\nb} \right) ^m \bin{m}{l} \sin ^2\left( \Omega _m\chi \right) ,\\
P_g^{c4} &=& \frac12 \sum _{j=0}^{\infty} e^{-2jT } \frac{\nb ^j - j\nb^{j-1}}{\left( 1+\nb \right) ^{j+1}} \sum _{l=0}^j \left( -\nb\right) ^{-l} \bin{j}{l} \sum _{m=l}^{\infty} \left( \frac{\nb}{1+\nb} \right) ^{m+1} \bin{m+1}{l} \cos ^2\left( \Omega _m\chi \right) ,
\end{eqnarray}
while the oscillatory term has the form
\begin{equation}
P_g^o = \frac{e^{-T}}2 \sum _{j=0}^{\infty} e^{-2jT} \frac{\nb ^j}{\left( 1+\nb \right) ^{j+2}} \sum _{l=0}^j \left( -\nb \right) ^l \bin{j}{l}\sum _{m=l}^{\infty} \left( \frac{\nb}{1+\nb} \right) ^m \bin{m}{l} \frac{\left( j+1\right) \left( m+1\right) ^{\frac12}}{l+1} \sin \left( 2\Omega _m\chi \right) .
\end{equation} 
The condition $\nb < 1$ is sufficient to make all the above series convergent.

The fringe pattern given by equation (\ref{pad}) has visibility
\begin{equation}
\calV = \frac{P_g^o}{P_g^c},
\end{equation}
and this visibility function is represented by the lower curve in figure 4. There we used $\nb = 0.7$, which characterizes the mean photon number of the mode when thermal equilibrium is attained, according to reference \cite{Beretal01}. It is important to note that the ``cooling'' process adopted in the experiment is important to make the initial state closer to vacuum, at least in the sense of mean photon number, but this process does not affect the reservoir temperature, which is the essential data here. For the time $T = .008$ used in the experiment, we get $\calV = 98.3\%$. Renormalized by the $\eta$ factor from the first setup, we get $\calV = 73.7\%$. As we can estimate the visibility of the experimental fringe pattern of figure 2 as $\calV = 69\%$, we can conclude that, in the actual experiment, cavity imperfections are small sources of visibility loss. In fact, with our calculations, we can conclude that the second setup would generate a $70.2\%$ visibility pattern if there were no cavity losses, but with all other experimental error sources present.

We now suggest a simple modification of this part of the experiment which would allow for monitoring the progressive appearance of classicality as follows. One can  enlarge the time in between the two $\frac{\pi}2$ pulses and monitor the visibility loss of the patterns. This can be achieved by slowing down the atoms, which actual experimental velocity in ref. \cite{Beretal01} is $500 m/s$. If one supposes that the remaining experimental errors will not vary, for a $200 m/s$ atomic beam one will obtain a $67\%$ visibility pattern; for $50 m/s$, $59\%$; and for $10 m/s$, $31\%$. It is interesting to interpret that in this case the classical manifestation that is being achieved is the loss of the capacity of making a ``quantum eraser''; in other words, the which way information initially available in the field mode, which could be manipulated, is being irreversibly lost to the environment. While in the first setup the atom-field entanglement is suppressed by the enlarging of field amplitude (large quantum numbers), in this second modified one this entanglement is being suppressed by the concurrent field-environment entanglement creation (decoherence).

It is instructive to compare this proposed experiment with a classical analog. In a purely optical Mach-Zehnder interferometer one can vary the size of the arms and progressively lose visibility up to the {\emph{coherence length}}. This marks a transition from the {\emph{wave optics}} domain, dominated by interference and diffraction effects, to {\emph{geometrical optics}}, characterized only by rays. In the proposed experiment one can progressively lose visibility in a Ramsey fringe pattern, up to a {\emph{decoherence time}}, marking the transition from {\emph{quantum mechanics}}, characterized by matter wave interference and diffraction effects, to {\emph{classical mechanics}}, characterized only by trajectories. One should note that the above optical-mechanical analogy has guided Schr\"odinger in proposing his wave equation for the ``new'' mechanics \cite{Sch26}, and our proposed experiment would follow this transition.

The authors would like to thank L. Davidovich, N. Zagury, R.L. de Matos Filho, J.G. Peixoto de Faria, and P. Nussenzveig for fruitful discussions on this matter. This work was partially supported by the Milleniun Institute on Quantum Information. MCN is partially supported by CNPq.

Figure captions

Figure 1: The Mach-Zehnder analog and the first experimental scheme. {\bf a}, The first beam splitter, $B_1$, is free to oscillate coupled to a spring fixed at $O$. If the $B_1$ mass is macroscopic, the momentum kick given by the reflected photon can be neglected and the usual interference pattern is observed. In the case where $B_1$'s mass is microscopic, $B_1$ movement would contain information about the photon trajectory, and the fringes would be eliminated. {\bf b}, A Rydberg atom internal eigenstate is taken to a superposition of two internal eigenstates by the interaction with the $C$ cavity field mode. If the field intensity is large, the effect on the field of interaction with the atom can be neglected, and Ramsey fringes are observed at the recombined ``beams'' atomic detection. In the low intensity field case, the mean photon number after the interaction stores information about the atomic state, and fringes are suppressed. The picture represents $e$ and $g$ levels by distinct trajectories just to improve scheme visualization. (Figure taken from reference \cite{Beretal01}).

Figure 2: The Mach-Zehnder analog, the second experimental scheme, and experimental results. {\bf a}, Both beam splitters, $B_1$ e $B_2$, are coupled to the same oscillator. Interaction with the second beam splitter, $B_2$, allows the ``erasure'' of the which path information created in the interaction with $B_1$, restoring interference fringes. {\bf b}, The two atom-field interacting regions $R_1$ e $R_2$ allows the ``erasure'' of which way information, restoring Ramsey fringes. {\bf c}, Experimental results are fitted by a $\calV = 69\%$ visibility interference pattern. (Figure taken from reference \cite{Beretal01}).

Figure 3: The modified Mach-Zehnder interferometer used to discuss a classical analogy to ``quantum'' eraser. {\bf a}, If the detector is aligned with one arm, only light from this arm is recorded, and no interference is obtained. {\bf b}, By imposing a linear combination of wave vectors one can obtain (or restore) a fringe pattern. In this sense, all interferometers work like an ``unconditional quantum eraser'', where the alignment process is made to make the alternative paths indistinguishable.

Figure 4: Pattern visibility as a function of the time between the two $\frac{\pi}2$ pulses. The upper curve is for null temperature, the lower for $\nb = 0.7$. Only thermal and dissipative effects are considered; the effect of all other imperfections is estimated in the text. $T$ is the adimensional time $T=k\tau$.


\begin{thebibliography}{99}
\bibitem{note} Our states $\ket{\alpha _e}$ and $\ket{\alpha _g}$ are not normalized, contrary to the ones in reference \cite{Beretal01}. This causes a discrepancy of a factor of $2$ between equivalent expressions. Besides this, we do not eliminate the accumulated phases, which collaborate to the observed shift in different fringe patterns.
\end{thebibliography}
\end{document}